\documentclass[aps,pra,twocolumn,groupedaddress]{revtex4-1}

\usepackage{amsmath}    
\usepackage{graphicx}   
\usepackage{verbatim}   
\usepackage{color}      
\usepackage{subfigure}  
\usepackage{hyperref}   
\newcommand{\beq}{\begin{equation}}
\newcommand{\eeq}{\end{equation}}
\newcommand{\bea}{\begin{eqnarray}}
\newcommand{\eea}{\end{eqnarray}}

\def\({\left(}
\def\){\right)}
\def\[{\left[}
\def\]{\right]}

\begin{document}


\title{Testing the photon-number statistics of a quantum key distribution light source}

\author{J. F. Dynes}
\author{M. Lucamarini}\email[]{marco.lucamarini@crl.toshiba.co.uk}
\author{K. A. Patel}
\author{A. W. Sharpe}
\author{Z. L. Yuan}
\author{A. J. Shields}

\affiliation{Toshiba Research Europe Ltd, 208 Cambridge Science Park, Cambridge CB4 0GZ, UK }

\date{\today}

\begin{abstract}
\noindent A commonly held tenet is that lasers well above threshold emit photons in a coherent state, which follow a Poissonian statistics when measured in photon number. This feature is often exploited to build quantum-based random number generators or to derive the secure key rate of quantum key distribution systems. Hence the photon number distribution of the light source can directly impact the randomness and the security distilled from such devices. Here, we propose a method  based on measuring correlation functions to experimentally characterise a light source's photon statistics and use it in the estimation of a quantum key distribution system's key rate. This promises to be a useful tool for the certification of quantum-related technologies.
\end{abstract}

\maketitle

\section{Introduction\label{sec:I}}
\noindent Quantum random number generators (QRNGs)~\cite{ROT94,YKI+99,MYC+16} and quantum key distribution (QKD)~\cite{BB84,Eke91,GRT+02} are the first quantum-related technology to leap out of the lab and reach the maturity necessary for the market.

The goal of a QRNG is to generate unpredictable numbers based on the laws of quantum physics. The typical example is a single, indivisible, photon impinging on a beam splitter~\cite{ROT94,JAW+00,SGG+00}, which ideally provides a uniformly distributed random bit, 0 or 1, depending on the output port it emerges from. By measuring the photons with a pair of photodetectors, it is possible to extract random strings that can be suitably post-processed and employed for cryptographic applications as well as for lotteries, gambling and scientific simulations. Other common implementations are based on events that are expected to follow a Poissonian distribution, for example the arrival time of photons emitted by a coherent light source~\cite{YKI+99,SR07,DYS+08,FWN+10,NZZ+14}. In other cases, as for QRNGs based on the phase noise of a laser~\cite{GTL+10,QCL+10,JCS+11,XQM+12,AAJ+14,YLD+14}, the Poissonian nature of the source can evidence the good functioning of the randomness-generating mechanism.

The aim of QKD, on the other hand, is to generate shared randomness between two distant parties, traditionally called Alice and Bob, through the exchange of quantum signals. The most common light source in QKD systems is a laser followed by an attenuator. The attenuation level depends on the particular protocol implemented, but in general it is set so that the intensity of the emitted pulses approaches the single-photon regime, which is captured by the following condition:
\begin{equation}
\mu=\sum_n p_n n < 1.
\label{eq:mu}
\end{equation}
In Eq.~\eqref{eq:mu}, $\mu$ is the mean photon number of each emitted pulse and $p_n$ is the probability to emit $n$ photons.
The calibration of the mean photon number is essential to guarantee the security of QKD systems. If $\mu$ is too large, the secret information is redundantly encoded in $n>1$ photons, allowing an eavesdropper (Eve) to access the secret information. Therefore $\mu$ has to be carefully set and it is typically monitored in real time in existing QKD systems~\cite{DDL+15}.

\begin{figure}
  \centering
  \includegraphics[width=0.75\columnwidth]{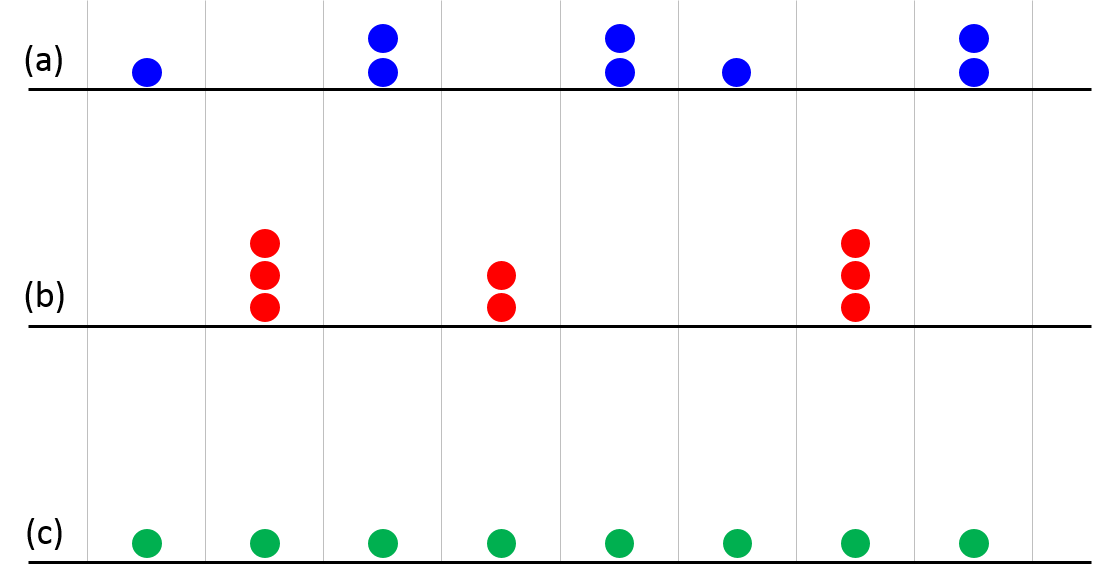}
  \caption{Schematic of photon number distributions in a train of optical pulses with average photon number $\mu=1$. (a) Poissonian source; (b) super-Poissonian source; (c) single-photon source. Each coloured dot represents a single photon. Vertical lines identify the optical pulses in the same time slot.}\label{fig:fig1}
\end{figure}

However, this is not the only security requirement and the statistics of the source, represented by $p_n$ in Eq.~\eqref{eq:mu}, also plays a crucial role. To gain some insight into this problem, consider Fig.~\ref{fig:fig1}, where three different types of photon number distributions are depicted, all displaying an average photon number $\mu=1$. Fig.~\ref{fig:fig1}(a) is the Poissonian case, with $p_n=e^{-\mu} \mu^n/n!$, which typically stems from a laser operated well above threshold~\cite{Lou00}. The resulting distribution is composed of statistically independent events and the number of photons in each pulse varies in a random way. In the super-Poissonian case, on the other hand, represented in Fig.~\ref{fig:fig1}(b), the light is composed of photons bunched in the same optical pulse. The typical example is a thermal source, which features a photon number statistics equal to $p_n=\mu^n/(1+\mu)^{1+n}$. Finally, we show in Fig.~\ref{fig:fig1}(c) the ideal case of a true single-photon source, which features $p_1=1$, $p_{n\neq1}=0$ and displays deterministic emission of exactly one photon per pulse.

Security-wise, the case (c) is the most secure, as there never is more than one photon carrying the secret information. On the contrary, the case (b) is the least secure, because the photon bunching effect favours the redundant encoding of the information.
This shows that even if the mean photon number $\mu$ is well characterised, the photon number distribution probability $p_n$ can still determine the insecurity of a QKD system. It is therefore important to find ways to characterise the photon statistics $p_n$ of a quantum-related light source.

For that, one method would be to use photon number resolving detectors or, equivalently, space-multiplexed or time-multiplexed threshold detectors~\cite{Had09}. However, in this case, a precise calibration of the detection efficiency would be required, which is far from trivial.
A second, more practical, method is the one proposed here. We perform a generalised Hanbury Brown-Twiss (HBT) experiment~\cite{HT56,HT57} to estimate the normalised correlation functions $g^{(m)}(\tau)$~\cite{Gla63,Gla63b} of the light source, with $m=\{2,3,4\}$. These functions, evaluated at time delay $\tau=0$, are closely related to the photon statistics of the light source. For example, it is well-known that if $p_n$ follows a Poissonian statistics, the light source has all the $g^{(m)}(0)$ equal to unity~\cite{Lou00}, irrespective of the order $m$. Although experimentally it is not possible to measure the correlation functions to all orders, we will show that measuring them up to the fourth order is sufficient to determine tight bounds on the secure key rate of a QKD system. To apply this method, we will work in the low detection efficiency approximation. This allows us to treat our threshold detectors as linear and makes our estimate independent of the exact efficiency of the detectors used in the experiment. The details of this approach are given in the next Sec.~\ref{sec:gn}. In Sec.~\ref{sec:III} we will describe the experimental implementation, whereas in Sec.~\ref{sec:theory} we will exploit our experimental results to estimate the secure key rate of a QKD system~\cite{LPD+13}.

\section{Normalised correlation functions\label{sec:gn}}

In the present work, we are interested in the characterisation of the photon statistics $p_n$ of a light source that is suitable for quantum-related technologies like a QRNG or a QKD system. We consider a scenario where the characterisation is worked out using detectors that are only loosely calibrated. In particular, we assume that the tester has guarantees that the efficiency of his detectors is positive but, at the same time, smaller than a certain threshold $\eta_0$. This allows us to treat threshold detectors as if they were linear and makes our result independent of the exact value of the detectors' efficiency. This assumption is perfectly reasonable if the source is tested in a trusted environment, where there is no eavesdropping ongoing during the test. It is then reasonable to assume that the tester has a certain control over his detectors and can arbitrarily decrease their efficiency below $\eta_0$ using a suitable attenuator.

Under this assumption, we can write the ``normalized correlation functions''~\cite{Gla63,Gla63b}, or ``degrees of coherence''~\cite{Lou00}, $g^{(2)}(0)$, $g^{(3)}(0)$, $g^{(4)}(0)$ of a stationary light source as:
\begin{align}
g^{(2)}(0)  &= \left\langle \hat{a}^{\dagger}\hat{a}^{\dagger}\hat{a}\hat{a}\right\rangle / \left\langle \hat{a}^{\dagger}\hat{a}\right\rangle^{2}\label{eq:g2}\\
g^{(3)}(0)  &= \left\langle \hat{a}^{\dagger}\hat{a}^{\dagger}\hat{a}^{\dagger}\hat{a}\hat{a}\hat{a}\right\rangle / \left\langle \hat{a}^{\dagger}\hat{a}\right\rangle^{3}\label{eq:g3}\\
g^{(4)}(0)  &= \left\langle \hat{a}^{\dagger}\hat{a}^{\dagger}\hat{a}^{\dagger}\hat{a}^{\dagger}\hat{a}\hat{a}\hat{a}\hat{a}\right\rangle / \left\langle \hat{a}^{\dagger}\hat{a}\right\rangle ^{4}\label{eq:g4}
\end{align}
where $\hat{a}^{\dagger}$ and $\hat{a}$ are the boson construction and destruction operators of a single quantum of light, respectively, and the bra-ket notation indicates the average operation over the states emitted by the source.

When the source is pulsed, which is often the case in QKD systems, the above relations can be written in a discrete form~\cite{SFV+04,SGN+14}:
\begin{align}
g^{(2)}[0]  &= \left\langle \hat{a}_k^{\dagger}\hat{a}_k^{\dagger}\hat{a}_k\hat{a}_k\right\rangle / \left\langle \hat{a}_k^{\dagger}\hat{a}_k\right\rangle^{2}\label{eq:g2d}\\
g^{(3)}[0]  &= \left\langle \hat{a}_k^{\dagger}\hat{a}_k^{\dagger}\hat{a}_k^{\dagger}\hat{a}_k\hat{a}_k\hat{a}_k\right\rangle / \left\langle \hat{a}_k^{\dagger}\hat{a}_k\right\rangle ^{3}\label{eq:g3d}\\
g^{(4)}[0]  &= \left\langle \hat{a}_k^{\dagger}\hat{a}_k^{\dagger}\hat{a}_k^{\dagger}\hat{a}_k^{\dagger}\hat{a}_k\hat{a}_k\hat{a}_k\hat{a}_k\right\rangle / \left\langle \hat{a}_k^{\dagger}\hat{a}_k\right\rangle ^{4}\label{eq:g4d}
\end{align}
where $k$ denotes the integer value representing the pulse number. In the following, we will use a compressed notation $g_m$ to indicate the quantity $g^{(m)}[0]$, dropping the zero time delay whenever it is unnecessary to specify it.

The normalized photon correlation functions in Eqs.~\eqref{eq:g2d}-\eqref{eq:g4d} represent the quantities measured in our experiment. Under the low detection efficiency approximation they are attractive from an experimental point of view as they can be measured using threshold single photon detectors. In Sec.~\ref{sec:III} we will describe the experiment performed to estimate these functions.

To connect the discrete correlation functions $g_{m}$ with the photon statistics $p_n$, we consider the density matrix of the state emitted by the source, expressed as a sum of photon number states:
\begin{equation}
\rho=\sum\nolimits_{n=0}^{\infty}p_n \left\vert n\right\rangle\left\langle n\right\vert.
\label{eq:ro}
\end{equation}
Upon rewriting Eqs.~\eqref{eq:g2d}-\eqref{eq:g4d} in terms of the density matrix in Eq.~\eqref{eq:ro}, and using the compressed notation previously introduced, we obtain:
\begin{align}
g_{2}  &  =\sum\nolimits_{n=0}^{\infty}p_n [n\left(n-1\right)] / \mu^2 \\
g_{3}  &  =\sum\nolimits_{n=0}^{\infty}p_n [n\left(n-1\right)\left(n-2\right)] / \mu^3 \\
g_{4}  &  =\sum\nolimits_{n=0}^{\infty}p_n [n\left(n-1\right)\left(n-2\right)\left(n-3\right)] / \mu^4
\end{align}
where $\mu$ has been defined in Eq.~\eqref{eq:mu}. From the above equations, it is apparent that the value of $\mu$ is necessary to find out the $n$-photon probability $p_n$. For example, it is intuitive that for large $\mu$ the $p_n$ will be centred around larger values of $n$. However, the exact determination of $\mu$ is not necessary for estimating the correlation functions $g_m$. Therefore we can treat $\mu$ as a parameter of the theory, known to the user Alice who prepares the states for the QKD protocol. It is worth noting that $\mu$ is usually determined quite easily and precisely in existing QKD systems by means of a calibrated power meter.

\section{Experimental setup and results\label{sec:III}}

\begin{figure}
  \centering
  \includegraphics[width=\columnwidth]{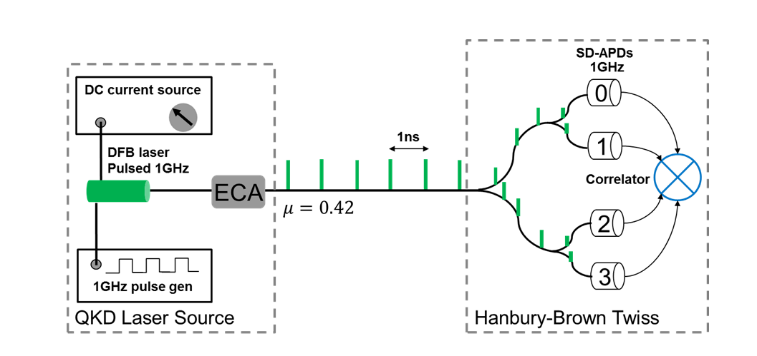}
  \caption{Schematic of the experimental arrangement for measuring normalized correlation functions up to the fourth order.  ECA: electronically controlled attenuator, SD-APD: self-differencing avalanche photodiode.}\label{fig:fig2}
\end{figure}

A schematic of the experimental arrangement used to characterise the $g_m$ functions is shown in Fig.~\ref{fig:fig2}. In order to connect our method with a prominent application, the light source to be characterised has been taken directly from a GHz-clocked QKD prototype~\cite{DDL+15}.

It consists of a DFB laser diode linked to a 1~ GHz pulse generator and driven by a DC current source. A small DC current is supplied to the laser diode from the DC current source, which causes a minor amount of spontaneous emission. The pulse generator produces a square wave with approximately 2V amplitude which modulates the DFB laser diode on top of the DC current. Under normal operating conditions, the modulation drives the laser diode in a gain switched mode, i.e., above and below lasing threshold. This causes a train of phase-randomised optical pulses to be produced at a frequency of 1 GHz~\cite{YLD+14}. The resulting optical pulses are then attenuated to the single photon level using an electronically controlled attenuator (ECA), so that the condition expressed by Eq.~\eqref{eq:mu} is satisfied in our experiments.

To analyse the correlation functions, a four channel HBT interferometer was constructed. Photons from the QKD laser source under test are transmitted into a 50:50 beamsplitter whereupon they encounter a second 50:50 beamsplitter.  Four self-differencing avalanche photodiodes (SD-APD)~\cite{YKS+07} synchronised to the QKD laser source detect the transmitted photons. Care was taken in the experiment to guarantee a detection efficiency well below 1\% in each detector. The subsequent electrical output from each SD-APD is then discriminated before being sent to a correlator card where time stamps are assigned to the photon arrivals.  The time stamps from the correlator were analysed by a custom built program.  The program constructed two, three and four-fold photon coincidences from the photon arrival times.  From this data, the resulting correlation functions up to the fourth order were evaluated.

\subsection{Dependence on laser drive current\label{subsec:drive-current}}

We start our analysis by examining the effect of changing the laser drive current on the normalised second order correlation function $g_2$.  This allows us to gain some insight on how the photon statistics might change for a laser diode under different operating conditions.

\begin{figure}
  \centering
  \includegraphics[width=0.95\columnwidth]{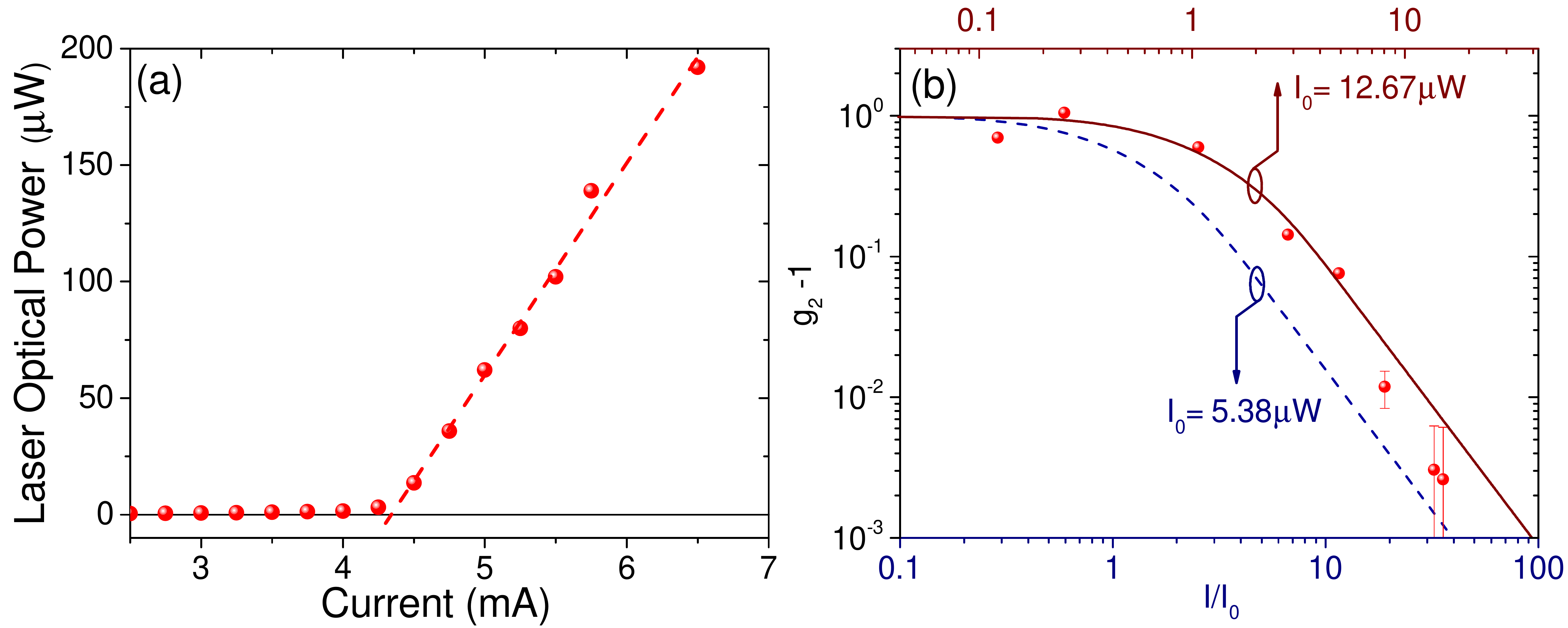}
  \caption{(a): Laser DC current vs optical output power from DFB laser diode. Above threshold, around $4.5$~mA, there is a steep rise in output power.  (b): $g_2 -1$ as a function of laser optical intensity $I$, normalised to threshold optical intensity $I_0$ with $I_0=12.67\mu$W. Points: Experimental data. Solid curve: fit of the non-linear oscillator model using $I_0$ as a free parameter. Dashed curve: non-linear oscillator model with $I_0=5.38\mu$W.}\label{fig:fig3}
\end{figure}

Fig.~\ref{fig:fig3}(a) shows the DC laser current vs optical output power dependence for the QKD laser source.  The 1 GHz square wave modulation is applied at all times.  Note the optical power is measured before the ECA using a standard optical power meter.  As expected, the optical output rises sharply after the lasing threshold, around $4.5$~mA, has been exceeded. Fig.~\ref{fig:fig3}(b) (points) shows the resulting normalised second order correlation function minus one ($g_2-1$) as measured in the single photon regime (i.e. after the ECA).  Below lasing threshold, $g_2-1 \approx 1$, indicating photon bunching, as might be expected for a thermal-like source. Around threshold, $g_2-1 $ starts to reduce.  As the DC current is further increased, the fluctuations begin to die out and $g_2-1$ tends to 0, indicating the laser emission is approaching Poissonian-like photon statistics, for which $g_2=1$.

To gain more insight into the functional dependence of $g_2-1$ on laser intensity, we fit the experimental data in Fig.~\ref{fig:fig3}(b) with a non-linear oscillator model~\cite{Man95}, using a single free parameter, the laser threshold intensity $I_0$. This model has been successful in describing $g_2$  around lasing threshold for gas lasers~\cite{AS66} provided the lasers under study can be approximated as single mode. The fit gives $I_0=12.67~\mu$W corresponding to a DC current of $4.49$~mA (solid curve in Fig.~\ref{fig:fig3}(b)). The fit appears reasonable for the experimental points around $I = I_0$ and up to $I = 6$. Beyond this laser intensity, the model overestimates $g_2-1$. Alternatively, we can extract a rough indication of lasing threshold $I_0$ from Fig.~\ref{fig:fig3}(a) whereby we fit the experimental data above $I_0$ using a straight line, yielding $I_0 = 5.38~\mu$W, corresponding to a DC current of $4.34$~mA. This gives the dashed curve in Fig.~\ref{fig:fig3}(b), now obtained with no free fitting parameters. In this case the experimental points in the range $I > I_0$ and $I < I_0 \approx 10$ are above the dashed curve although the points $ > I_0 \approx 10$ are closer to the prediction of this model.

The above analysis suggests the underlying model of the non-linear oscillator qualitatively describes the experimental data, if not quantitatively. Most likely the  ideal model lies between the two curves in Fig.~\ref{fig:fig3}(b). This indicates the excess $g_2-1$ for the experimental points with the three highest intensities is due to additional fluctuations on top of the noise predicted by the model. Note, as we show below, we do not rely on any physical model to construct bounds on the secure bit rate, see Sec.~\ref{sec:theory}.

For use in QKD, it is important to choose an appropriate DC current level.  Too low gives large photon bunching and therefore a high incidence of multi-photon events, which greatly facilitates photon number splitting attacks by an eavesdropper.  Too high a DC current yields Poissonian behaviour but increases the laser CW optical background.  A high CW optical background increases the phase correlation between successive optical pulses and destroys the phase randomisation of the QKD laser source~\cite{YLD+14}.

We choose a compromise such that under normal operating conditions, i.e. during QKD, the DC current is set to approximately 6.5~mA which corresponds to $g_2-1$ of the order of $10^{-3}$, as shown in Fig.~\ref{fig:fig3}(b).

Next we characterise higher order correlations.  By way of comparison higher order correlations are measured for two cases.  Firstly, below lasing threshold at a DC current of about 4.0~mA.  Secondly, above lasing threshold, where the DC current is 6.5~mA corresponding to normal operating conditions.

\subsubsection{Variation of  $g_m$ for laser below threshold\label{subsubsec:below-thresh}}

We first examine the variation of the higher order correlation functions when the laser is operated below threshold.  A DC current of 4.0~ mA is selected.

\begin{figure}
  \centering
  \includegraphics[width=0.95\columnwidth]{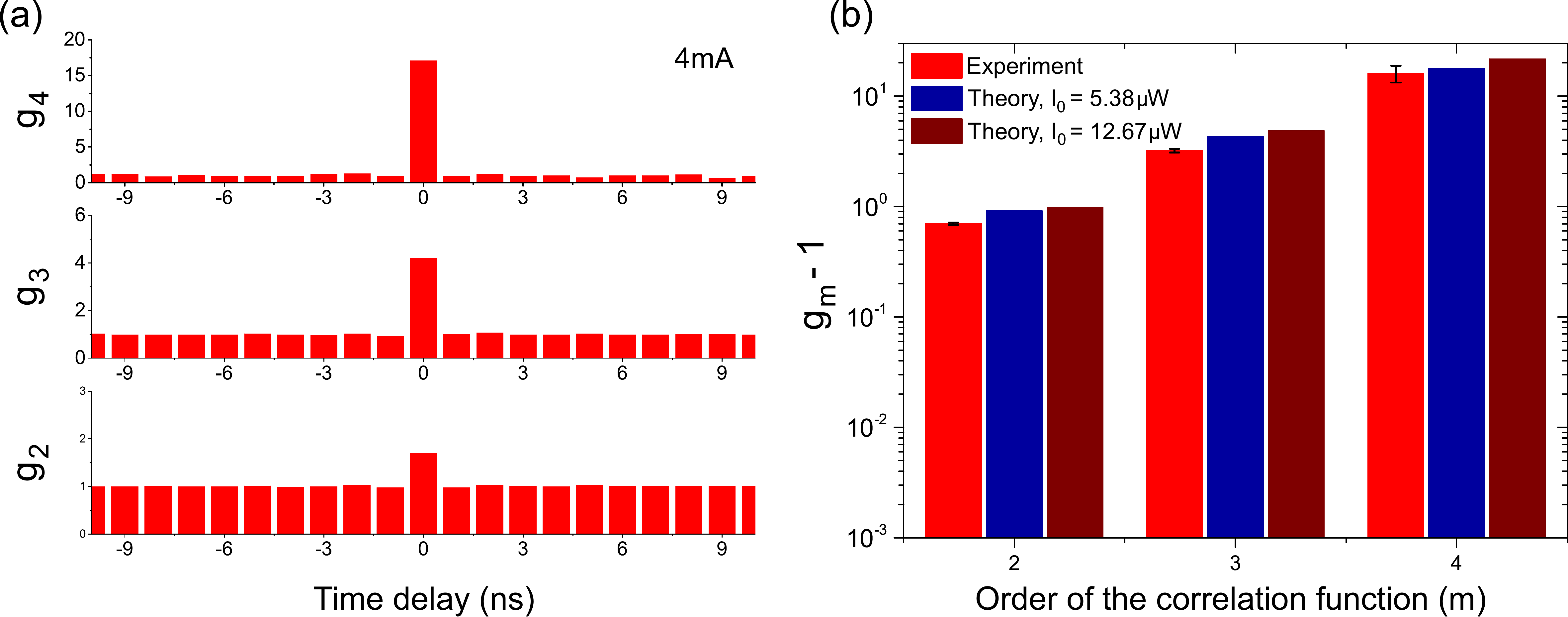}
  \caption{(a): Measured normalised correlation functions $g_2$, $g_3$ and $g_4$ for a DC current of 4.0~mA.  (b): $m$-th order $g_m -1$ as a function of the order $m$. Red bars: experimental data; blue bars: prediction of non-linear oscillator model when $I_0=5.38~\mu$W; wine bars: prediction of non-linear oscillator model when $I_0=12.67~\mu$W.}\label{fig:fig4}
\end{figure}

Fig.~\ref{fig:fig4}(a) shows the measured normalised correlation functions $g_2$, $g_3$ and $g_4$ and the numerical values are listed in the following equations:
\begin{eqnarray}
  g_2 &=& 1.6985 \pm 0.0138, \label{eq:constr1t}\\
  g_3 &=& 4.21   \pm 0.13,   \label{eq:constr2t}\\
  g_4 &=& 17.11  \pm 2.84.   \label{eq:constr3t}
\end{eqnarray}
The source displays a high level of bunching, approaching a thermal distribution, hence we term it ``quasi-thermal''. The peak at time zero grows with the order $m$, as shown with red bars in Fig.~\ref{fig:fig4}(b). Also plotted are the predictions of the non-linear oscillator model for the two laser threshold intensities shown in Fig.~\ref{fig:fig3}(b). Both scenarios predict slightly stronger correlations for all orders compared to what was experimentally measured. This behaviour can be intuitively understood by assuming the source is mainly bunched but also possesses a smaller Poissonian component since the measured $g_2=1.6985$ is less than the value $g_2=2$ expected for a purely thermal source. The bunching contribution naturally gives stronger correlation values $g_m$ as $m$ is increased but at the same time the smaller Poissonian component contributes to reducing these correlation values.

\subsubsection{Variation of  $g_m$ for laser above threshold\label{subsubsec:above-thresh}}

We now turn our attention to the variation of the higher order correlation functions when the laser is operated above threshold.  A DC current of ∼ 6.5~mA is selected for this purpose.
\begin{figure}
  \centering
  \includegraphics[width=0.95\columnwidth]{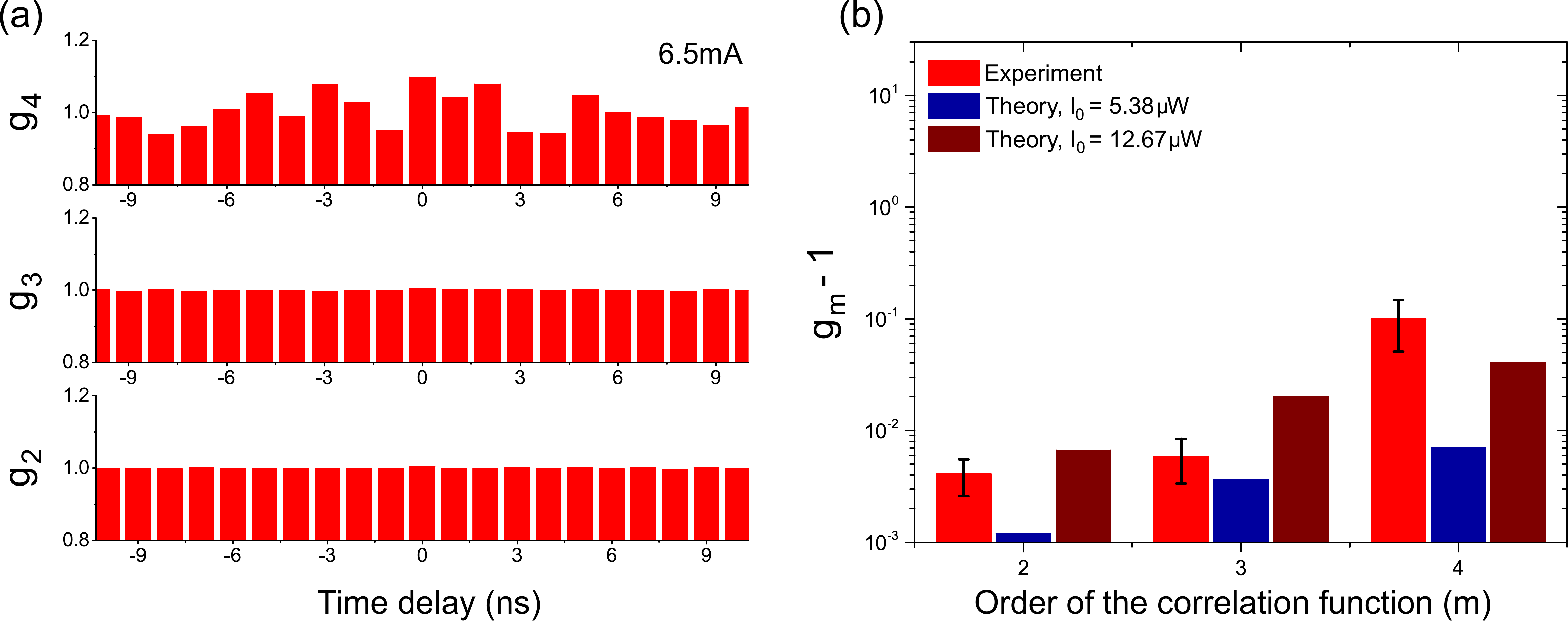}
  \caption{(a): Measured normalised correlation functions $g_2$, $g_3$ and $g_4$ for a DC current of 6.5~ mA. (b): $n$-th order normalised correlation functions as a function of the order $n$. Red bars: experimental data; blue bars: prediction of non-linear oscillator model when $I_0=5.38~\mu$W; wine bars: prediction of non-linear oscillator model when $I_0=12.67~\mu$W.}\label{fig:fig5}
\end{figure}
Fig.~\ref{fig:fig5}(a) shows the measured normalised correlation functions $g_2$, $g_3$ and $g_4$ and the numerical values are:
\begin{eqnarray}
  g_2 &=& 1.0041 \pm 0.0039, \label{eq:constr1}\\
  g_3 &=& 1.0059 \pm 0.0056, \label{eq:constr2}\\
  g_4 &=& 1.099 \pm 0.049.  \label{eq:constr3}
\end{eqnarray}
This time there is no clear peak at time zero, which indicates the laser is qualitatively Poissonian.  However, plotting the $g_m$ as a function of $m$ reveals an ever so slight increase in correlation despite the large error bars, Fig.~\ref{fig:fig5}(b), red bars. On this occasion the source appears to have a minor bunched-like component, hence much weaker correlation values $g_m$ are observed as the order $m$ is increased, compared to the previous case.

Again we plot the predictions of the non-linear oscillator model for the two laser threshold intensities shown in Fig.~\ref{fig:fig3}(b). For $g_2$ and $g_3$ the measured data is between the two predictions. For $g_4$ both predictions slightly underestimate $g_4$. The reason for this is unclear but the error bar for the $g_4$ measured point is almost an order of magnitude greater than the error bars for $g_2$ and $g_3$. This stems from the four-fold coincidence count rate being considerably lower than three or two-fold coincidence count rates. Furthermore, $g_4-1$ is quite small, $< 0.1$. This can be compared with $g_4-1$ in Fig.~\ref{fig:fig4}(a), which is around a factor of hundred larger.

Note for the measured correlation functions we have also considered corrections from afterpulsing, dark counts and dead time. However these corrections are well within the size of one error bar, so have consequently been safely neglected in the current analysis.

\section{Worst-case bounds from correlation functions and secure key rate\label{sec:theory}}

Let us now consider how the experimental results from the previous section can be used to determine the secure key rate of a QKD system. For that, we rewrite the key rate equation for the efficient~\cite{LCA05,SR08} decoy-state~\cite{Hwa03,Wan05,LMC05} BB84 protocol~\cite{BB84} in the finite-size scenario~\cite{HN14,LPD+13,LCW+14} in the following way:
\begin{equation}
  R \geq p_u p_Z^2 \{ \underline{p}_1 \underline{y}_{1,Z} [1-h(\overline{e}_{1,X})]- f Q_{Z}h(E_Z)-\Delta \}.
  \label{eq:rate}
\end{equation}
With minor modifications, Eq.~\eqref{eq:rate} can be easily adapted to other BB84-like protocols. To write it, we considered a protocol where Alice prepares phase-randomised weak coherent states in the two bases $Z$ and $X$, randomly chosen with probabilities $p_Z$ and $p_X=1-p_Z$, respectively, and in the three intensity classes $u$ (signal), $v$ (decoy) and $w$ (vacuum), randomly chosen with probabilities $p_w$, $p_v$ and $p_u=1-p_v-p_w$, respectively. However, the users extract the secure key bits only from the basis $Z$ and from the class $u$.
The quantity $R$ is the fraction of secure bits per signal distilled by the system; when multiplied by the system's clock rate, it provides the amount of secure bits in the time unit; $\overline{e}_{1,X}$ is the upper bound to the single-photon QBER in the $X$ basis and $\underline{y}_{1,Z}$ is the lower bound to the 1-photon yield in the $Z$ basis. This last quantity is multiplied by the lower bound to the 1-photon probability, $\underline{p}_1$, thus providing an overall lower bound to the 1-photon gain of the protocol. Finally, $f$ is the inefficiency of error correction, $Q_{Z}$ and $E_Z$ are the gain and the QBER, respectively, measured in the $Z$ basis and $\Delta$ is a parameter related to the finite-size effect.

The crucial difference between the rate in Eq.~\eqref{eq:rate} and previous QKD key rates resides in the term $\underline{p}_1$. Usually, a Poissonian distribution is assumed for decoy-state QKD, entailing that $p_1=\mu e^{-\mu}$~\cite{Hwa03,Wan05,LMC05}. However, by applying our method, we can now drop this assumption and replace it with the lower bound $\underline{p}_1$ that is measured in the experimental characterisation of the light source. Similar bounds can be obtained for the other probabilities $p_0$, $p_{\geq2}$, as well as for the other parameters $y_0$, $y_1$, $e_1$ usually present in decoy-state QKD. The details on the constrained numerical optimization providing such bounds are given in the Appendix.

\begin{figure}[tpb]
  \centering
  \includegraphics[width=\columnwidth]{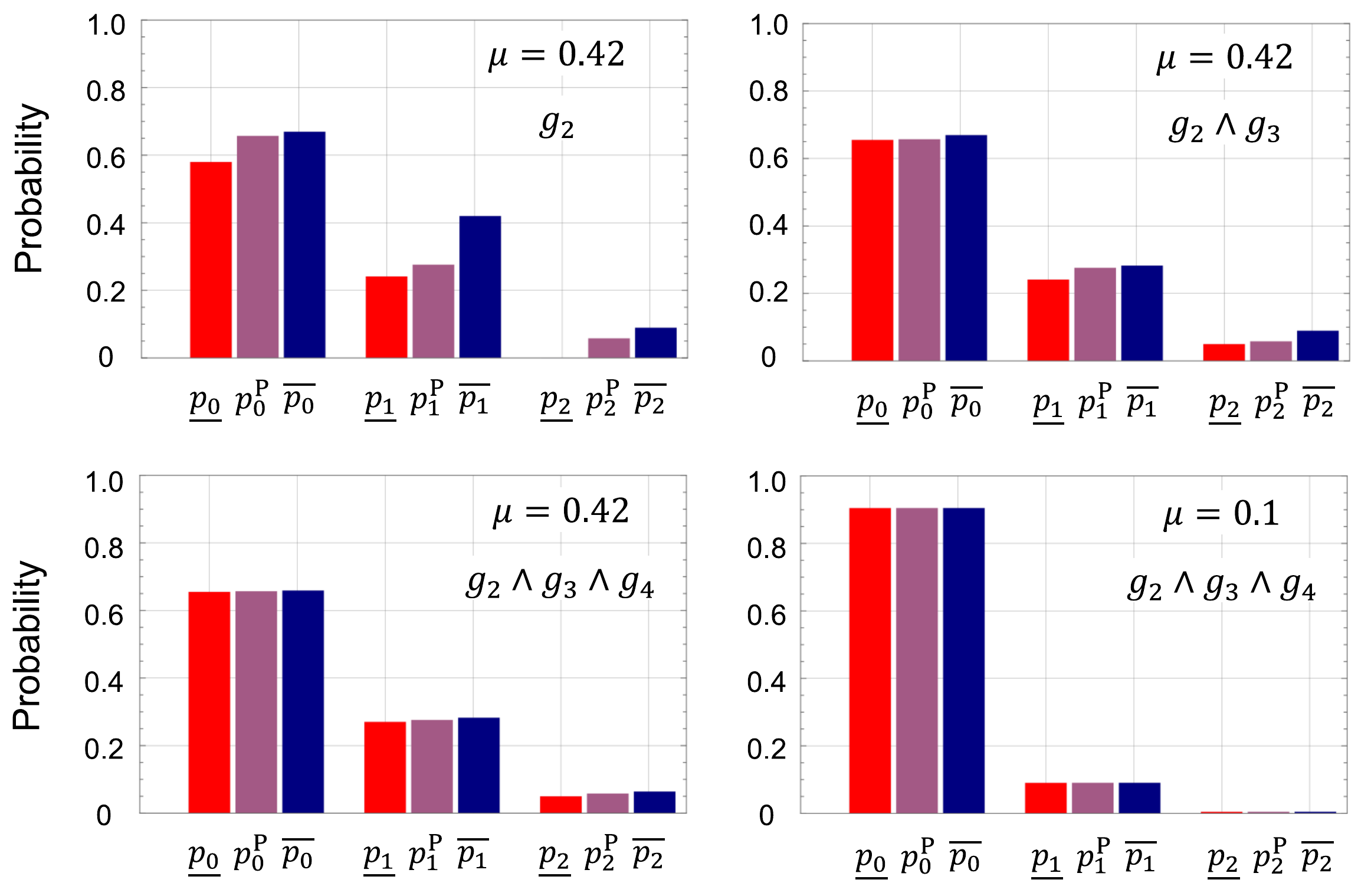}
  \caption{Bounds to the photon probabilities $p_0$, $p_1$, $p_2$ for various experimental conditions. In the top right corner of each diagram, the correlation functions used in the estimation of the bounds and the mean photon number $\mu$ are indicated. Underlined (overlined) quantities are for lower (upper) bounds obtained from the experimental data given in Sec.~\ref{subsubsec:above-thresh}, Eqs.~\eqref{eq:constr1}-\eqref{eq:constr3}, for a confidence interval of 7 standard deviations. The superscript P refers to the ideal Poissonian distribution.
  \underline{Numerical Values}. Poissonian: $p_n^{\textrm{P}}=e^{-\mu}\mu^n/n!$, with $\mu=\{0.42,0.1\}$. Bounds -- clockwise starting from top-left diagram:	
  $\underline{p_0} =$ $\{0.580; $ $	0.655;   $ $0.905;   $ $0.655;   \}$
  $\overline{p_0}  =$ $\{0.670  ; $ $0.669;   $ $0.905;   $ $0.659;   \}$	
  $\underline{p_1} =$ $\{0.2411 ; $ $0.2411;  $ $0.0903;  $ $0.2702;  \}$
  $\overline{p_1}  =$ $\{0.4203 ; $ $0.2826;  $ $0.0906;  $ $0.2826;  \}$	
  $\underline{p_2} =$ $\{0     ; $ $0.04973; $ $0.00446; $ $0.04973; \}$
  $\overline{p_2}  =$ $\{0.08947; $ $0.08947; $ $0.00461; $ $0.06405. \}$
  }\label{fig:fig6}
\end{figure}

\begin{figure}[tpb]
  \centering
  \includegraphics[width=\columnwidth]{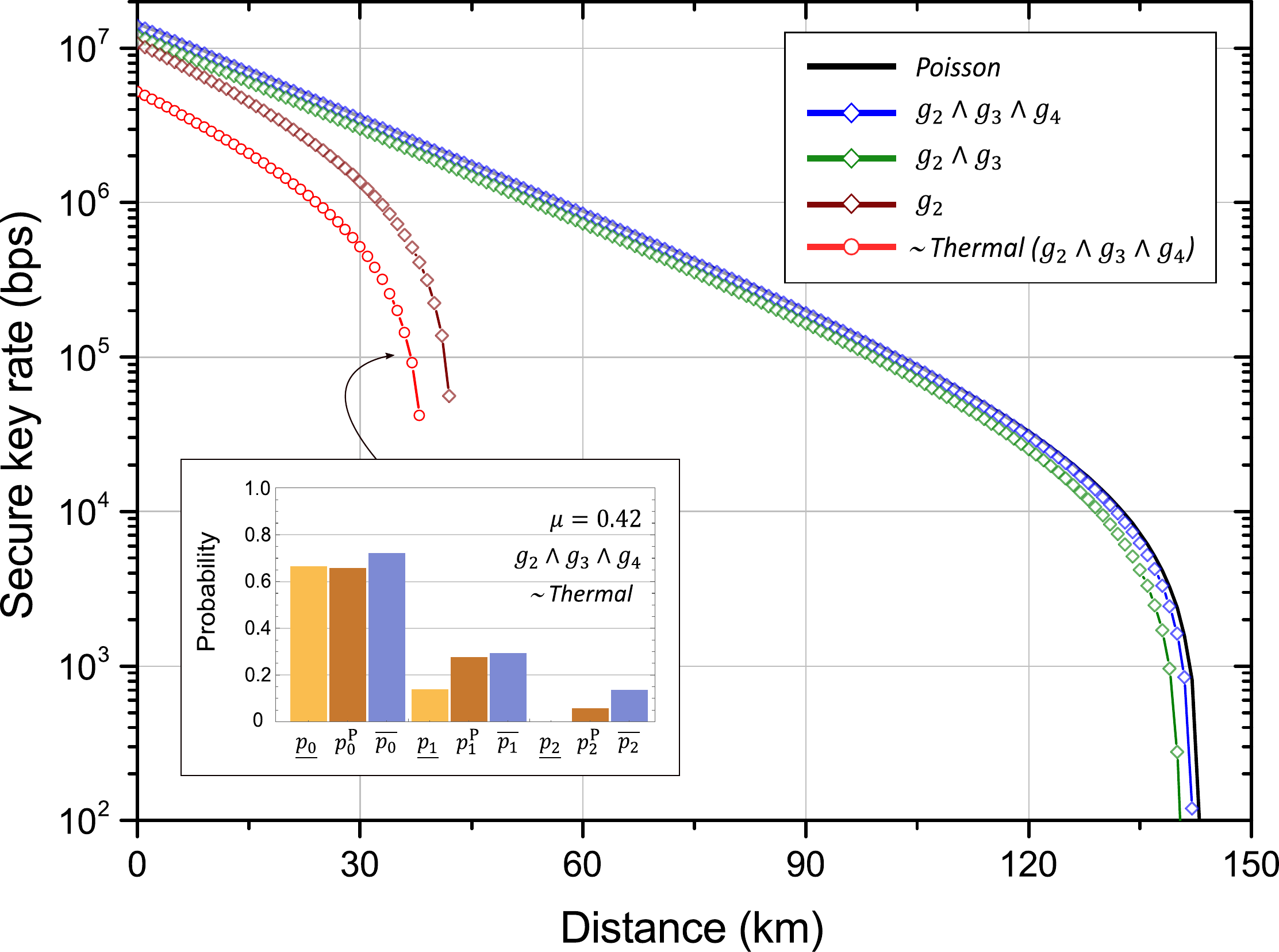}
  \caption{Secure key rate of the efficient decoy-state BB84 protocol in the finite-size scenario, for the ideal case of a Poissonian source (black line) and for the real case where the correlation functions up to the fourth order have been measured.
  The values of the correlation functions have all been drawn from the experiments. The blue, green and wine lines correspond to the source described in Sec.~\ref{subsubsec:above-thresh}, Eqs.~\eqref{eq:constr1}-\eqref{eq:constr3}, whereas the red line is for the source in Sec.~\ref{subsubsec:below-thresh}, Eqs.~\eqref{eq:constr1t}--\eqref{eq:constr3t}. The bounds for this latter source, which displays a quasi-thermal distribution, are given in the inset, together with the Poisson probabilities for comparative purposes. The mean photon number for the signal and the decoy states in the protocol have been set to $u=0.42$ and $v=0.02$, respectively, and the security parameter is $\epsilon<10^{-10}$. }\label{fig:fig7}
\end{figure}

As an example, in Fig.~\ref{fig:fig6} we provide the bounds for $p_0$, $p_1$ and $p_2$ drawn from the experimental values reported in Eqs.~\eqref{eq:constr1}-\eqref{eq:constr3} and we compare them with the corresponding probabilities from the Poissonian distributions, for various experimental conditions. On the top-left diagram, we consider the case where Alice prepares a mean photon number $\mu=0.42$, a common value for the signal state in decoy-state QKD, and uses only the $g_2$ function from Eq.~\eqref{eq:constr1} as a constraint. In this case, the comparison with the ideal Poissonian probabilities normally used in decoy-state QKD shows that the bounds are quite loose. Therefore, we expect a dramatic reduction of the key rate obtained by assuming that the distribution is Poissonian. However, if more constraints are added, the bounds become tighter and tighter, as can be noted from the top-right and bottom-left diagrams in Fig.~\ref{fig:fig6}. This should entail a close-to-ideal key rate in these cases. Finally, when the mean photon number is decreased to $\mu=0.1$, a value commonly used for decoy states in QKD, and all the constraints are applied, the match with the Poissonian probabilities is nearly perfect (see bottom-right diagram in Fig.~\ref{fig:fig6}), implying a negligible reduction in the key rate.

In the estimation leading to Fig.~\ref{fig:fig6}, we have considered that the experimental errors follow a Gaussian distribution and we have taken $\gamma=7$ standard deviations to define the related confidence interval. This corresponds to a probability $\epsilon=2.56\times10^{-12}$ that the true experimental value falls outside the confidence interval. In turn, once all the confidence intervals related to the constrained optimisation are taken into account~\cite{LDY+12}, this probability generates a global security parameter for the QKD protocol in the finite-size scenario~\cite{MQZ+05,ZZR+17} well below $10^{-10}$.

As a prominent application of our method, we use the bounds just derived to estimate the secure key rate of a QKD system according to Eq.~\eqref{eq:rate}. The results are shown in Fig.~\ref{fig:fig7}. It can be seen that when the $g_2$ function is the only constraint in the problem (wine-coloured line), there is a considerable decrease in key rate and maximum distance achieved by protocol. On the contrary, when also the $g_3$ and $g_4$ are taken into account (green and blue lines), the key rate is very close to the ideal one. Intuitively, this can be explained by the fact that each additional constraint improves the bounds on the $p_n$. When only $g_2$ is considered, we can only reliably bound the first term in the photon statistics, $p_0$, which does not enter the key rate equation \eqref{eq:rate}, or it plays only a minor role in other versions of the BB84 protocol~\cite{LCW+14}. However, when both $g_2$ and $g_3$ are considered, we can bound $p_0$ and $p_1$, the latter being the most important term in the secure key rate. This explains the large gap in the plotted key rates between the orange line ($g_2$ only) and the red line ($g_2$ and $g_3$ together).

In the same figure, we also plot in red colour the key rate corresponding to the quasi-thermal source described in Sec.~\ref{subsubsec:below-thresh}, which is obtained by plugging Eqs.~\eqref{eq:constr1t}--\eqref{eq:constr3t} into the corresponding optimisation problem. The upper and lower bounds are shown in the inset of Fig.~\ref{fig:fig7}. It can be noticed that the lower bound for the zero-photon fraction of this distribution (leftmost yellow bar) is higher than the corresponding Poisson probability (brown bar), clearly showing that the distribution is not Poissonian. Nevertheless, it is still possible to extract from it a key rate that remains positive up to a distance of about 40~km in optical fibre. Larger distances might be achieved upon optimisation of the system's parameters. This positive key rate is remarkable given the highly bunched nature of the source and represents an experimental evidence of a positive key rate drawn from a quasi-thermal source.

On the other hand, this result should be treated as a caveat for the correct operation of a laser source. If the laser is unawarely operated close-to-threshold, the users will not realise that the correct key rate is the one corresponding to a quasi-thermal source (red line in Fig.~\ref{fig:fig7}) and will keep using the privacy amplification corresponding to a Poissonian source. This entails that less privacy amplification than necessary will be performed, compromising the security of the system. The calibration of the QKD system's light source and the detailed description of its operative parameters represent a desirable best practice to prevent these kind of security risks.

\section{Conclusion\label{sec:concl}}

In quantum-related technology, claims about randomness and security crucially depend on a set of assumptions that have to be carefully met in the implementation. In the present work, we introduced and experimentally realised a test for one of the most common assumptions in QRNG devices and QKD systems, i.e., the Poissonian nature of the light source.

The method is based on a generalised HBT experiment~\cite{HT56} with single-photon detectors in the low-efficiency approximation. If the detection efficiency is known to be small, the exact efficiency of each detector simplifies in the corresponding expressions of the normalised correlation functions, Eqs.~\eqref{eq:g2d}-\eqref{eq:g4d}. This lets us bypass an accurate calibration of the detectors, which is often a cumbersome process.

On the other hand, our method makes it possible to test a light source upon which the tester has some limited prior knowledge. For example, we showed that the source in this work qualitatively follows a well-established laser model. However, the technique does not make use of this model and makes no stringent assumptions on the physics  underlying the photon source under test. This identifies a practical way to undertake calibration tests, which are normally performed in the protected environment of a laboratory, where the assumptions on the potential presence of a malevolent agent are mild.

The results of the experimental tests, given in Eqs.~\eqref{eq:constr1t}-\eqref{eq:constr3t} for a laser operated close to threshold and in Eqs.~\eqref{eq:constr1}-\eqref{eq:constr3} for a laser above threshold, show that photon statistics of the light source crucially depend on the driving current.
When the laser is operated above threshold, the photon distribution is very close to Poissonian, which is especially true when the mean photon number is low. In this case, the normalised correlation functions up to the fourth order are all close to 1. This implies that the bounds to the Poissonian distribution, depicted in Fig.~\ref{fig:fig6}, are quite tight. Nevertheless, when plugged in a typical key rate equation for a QKD system, these bounds generate a key rate that can deviate from the ideal one. This is shown in Fig.~\ref{fig:fig7}, which suggests that measuring at least the $g_2$ and $g_3$ correlation functions is essential to avoid a major reduction of the secure key rate.
This is feasible and practical, as we have shown, thus promoting the present method as a useful tool to guarantee the security of QKD systems.
Close to threshold, the distribution of the emitted photons becomes quasi-thermal and displays a highly bunched behaviour. However, even without optimising the system's parameters, we could obtain a positive key rate over a few tens of kilometres of optical fibre.

\begin{acknowledgments}
During the completion of this work, a paper on a closely-related subject appeared in~\cite{KSK17}.
We acknowledge useful discussions with Toshihiro Sasaki. The present work was supported by the project EMPIR 14IND05 MIQC2. This project has received funding from the EMPIR programme co-financed by the Participating States and from the European Unions Horizon 2020 research and innovation programme.
\end{acknowledgments}

\appendix

\begin{widetext}

\section{Bounds for the photon statistics}
\label{sec:appA}

In this section we show how to bound the photon probabilities $p_n$ using the experimentally measured correlation functions $g_m$, with $m=\{2,3,4\}$. This is a constrained optimisation problem that can be cast in linear form, so that it is efficient and provides a global solution. In this problem, the probabilities $p_n$ are the objective functions and the experimental correlation functions are the constraints.

We will consider in the following a specific problem, i.e., the minimisation of the single-photon probability $p_1$ when Alice emits optical pulses with average photon number $\mu$. Other problems related to the maximisation of $p_1$, or to the optimisation of the other $p_n$'s, with $n \neq 1$, are analogous and can be straightforwardly obtained from the one described here.

\subsection*{Asymptotic case}

We initially analyse the problem in the so-called asymptotic scenario, where it is assumed that an infinitely large data sample is available to the users. In the next sections we will generalise this case to the finite-size setting. The optimisation problem we consider can be expressed as:
%
\begin{center}
\renewcommand{\arraystretch}{1.5}
\begin{tabular}[c]{rrl}
min: & $\hat{p}_{1}$ & \\
subject to: 
& $1$ & $={\sum\nolimits_{n=0}^{\infty}}~\hat{p}_{n}$\\
& $\hat{g}_{2}$ & $={\sum\nolimits_{n=0}^{\infty}}~\hat{p}_{n}n\left( n-1\right)  /{\mu}^{2}$\\
& $\hat{g}_{3}$ & $={\sum\nolimits_{n=0}^{\infty}}~\hat{p}_{n}n\left( n-1\right)  \left( n-2\right)  /{\mu}^{3}$\\
& $\hat{g}_{4}$ & $={\sum\nolimits_{n=0}^{\infty}}~\hat{p}_{n}n\left( n-1\right)  \left( n-2\right)  \left( n-3\right)/{\mu}^{4}$\\
& $0$ & $\leq\hat{p}_{n}\leq1$\\
& $\underline{g_{2}}$ & $\leq\hat{g}_{2}\leq\overline{g_{2}}$\\
& $\underline{g_{3}}$ & $\leq\hat{g}_{3}\leq\overline{g_{3}}$\\
& $\underline{g_{4}}$ & $\leq\hat{g}_{4}\leq\overline{g_{4}}$
\end{tabular}
\begin{equation}\label{ref_problem}\end{equation}
\end{center}
where we have used hatted letters to indicate the variables of the problem and underlined or over-lined letters to indicate lower or upper bounds to the corresponding physical quantities, respectively. These bounds are specified as follows:
\begin{align}
\underline{g_{2}} &  =g_{2}-\gamma\times{\sigma_{g_{2}}} ~;~~ \overline{g_{2}}=g_{2}+\gamma\times\sigma_{g_{2}}  \\
\underline{g_{3}} &  =g_{3}-\gamma\times{\sigma_{g_{3}}} ~;~~ \overline{g_{3}}=g_{3}+\gamma\times\sigma_{g_{3}}  \\
\underline{g_{4}} &  =g_{4}-\gamma\times{\sigma_{g_{4}}} ~;~~ \overline{g_{4}}=g_{4}+\gamma\times\sigma_{g_{4}}
\end{align}
where the $\sigma$ quantities account for the experimental errors and $\gamma$ specifies the number of standard deviations defining the confidence interval related to the experiment, assuming a Gaussian distribution of the associated random variable. The problem in Eq.~\eqref{ref_problem} is linear in the unknowns and can be readily solved to find the global minimum for $\hat{p}_1$.

\subsection*{Finite-size case}

The problem in Eq.~\eqref{ref_problem} contains sums that run on an infinite number of terms. This is impractical to treat, so we turn it into a problem that contains a finite number of terms. The idea behind this is that when Eq.~\eqref{eq:mu} is fulfilled, only the first few terms in the sums give a relevant contribution. To write the finite-size optimisation problem, we will limit the number of photons $n$ to the interval $[0, N_{\textrm{cut}}]$, with $N_{\textrm{cut}}$ a natural number chosen large enough to satisfy the following condition:
\begin{equation}\label{eq:condition1}
  \sum_{n=0}^{N_{\rm cut}} p_n \geq \sum_{n=N_{\rm cut}}^{\infty} p_n .
\end{equation}
By choosing $N_{\rm cut}$ large enough, the condition in Eq.~\eqref{eq:condition1} can always be fulfilled. On the practical side, we choose $N_{\rm cut}=25$ in our solution of the optimisation problem. This entails that we are discarding all the photon distributions where the probability of having more than 25 photons in a single pulse is larger than the probability of having less than 25 photons in a single pulse. We believe that this assumption is met in all the distributions of practical interest. For example, if we consider a source where the measured average photon number $\mu$ is smaller than one, as in Eq.~\eqref{eq:mu}, our solution rules out a distribution where one optical pulse contains 25 photons and the following 24 pulses are empty, whereas it includes the three distributions depicted in Fig.~\ref{fig:fig1}. Moreover, also the most exotic photon distributions could be covered by further increasing the value of $N_{\rm cut}$, which is feasible, due to the linearity of the problem.

By using the closure condition $\sum_{n=0}^{\infty} p_n = 1$ we can rewrite the assumption in Eq.~\eqref{eq:condition1} as $\sum_{n=0}^{N_{\rm cut}} p_n \geq 1/2$. This, in turn, provides the following bounds for the sum of the variables $\hat{p}_n$'s in the finite-size scenario:
\begin{equation}\label{eq:bound1}
  1 \geq\sum\nolimits_{n=0}^{N_{\textrm{cut}}}\hat{p}_{n}\geq\frac{1}{2}.
\end{equation}
The lower bound in the above equation can be used to rewrite the lower bounds for the other constraints of the optimisation problem:
\begin{align}
&\sum\nolimits_{n=0}^{N_{\textrm{cut}}}\hat{p}_{n} n   ~\geq {\mu}/2  , \label{eq:lowbounds1}\\
&\sum\nolimits_{n=0}^{N_{\textrm{cut}}}\hat{p}_{n}n^{2} \geq ({\mu}^{2}\hat{g}_{2}+{\mu})/2  ,\label{eq:lowbounds2}\\
&\sum\nolimits_{n=0}^{N_{\textrm{cut}}}\hat{p}_{n}n^{3} \geq ({\mu}^{3}\hat{g}_{3}+3{\mu}^{2}\hat{g}_{2}+{\mu})/2  ,\label{eq:lowbounds3}\\
&\sum\nolimits_{n=0}^{N_{\textrm{cut}}}\hat{p}_{n}n^{4} \geq ({\mu}^{4}\hat{g}_{4}+6{\mu}^{3}\hat{g}_{3}+7{\mu}^{2}\hat{g}_{2}+{\mu})/2 \label{eq:lowbounds4}.
\end{align}
To upper bound the finite sum related to the average photon number, we exploit the following chain of relations:
\begin{align}
{\mu} &=\sum\nolimits_{n=0}^{\infty}\hat{p}_{n}n\\
    & = \sum\nolimits_{n=0}^{N_{\textrm{cut}}}\hat{p}_{n}n+\sum\nolimits_{n=N_{\textrm{cut}}+1}^{\infty}\hat{p}_{n}n\\
    & \geq\sum\nolimits_{n=0}^{N_{\textrm{cut}}}\hat{p}_{n}n+\left(N_{\textrm{cut}}+1\right) \sum\nolimits_{n=N_{\textrm{cut}}+1}^{\infty}\hat{p}_{n}.
\end{align}
Similar relations can be easily found for the correlation functions. The resulting upper bounds for the constraints, paired with those in Eqs.~\eqref{eq:lowbounds1}-\eqref{eq:lowbounds4}, are:
\begin{align}
{\mu}-\left(  N_{\textrm{cut}}+1\right)  \left(  1-\sum\nolimits_{n=0}^{N_{\textrm{cut}}}\hat{p}_{n}\right)   &  \geq\sum\nolimits_{n=0}^{N_{\textrm{cut}}}\hat{p}_{n}n\\
{\mu}^{2}\hat{g}_{2}+{\mu}-\left(  N_{\textrm{cut}}+1\right)  \left(  {\mu}-\sum\nolimits_{n=0}^{N_{\textrm{cut}}}\hat{p}_{n}n\right)   &  \geq\sum\nolimits_{n=0}^{N_{\textrm{cut}}}\hat{p}_{n}n^{2}\\
{\mu}^{3}\hat{g}_{3}+3{\mu}^{2}\hat{g}_{2}+{\mu}-\left(  N_{\textrm{cut}}+1\right)  \left(  {\mu}^{2}\hat{g}_{2}+{\mu}-\sum\nolimits_{n=0}^{N_{\textrm{cut}}}\hat{p}_{n}n^{2}\right)   &  \geq\sum\nolimits_{n=0}^{N_{\textrm{cut}}}\hat{p}_{n}n^{3}\\
{\mu}^{4}\hat{g}_{4}+6{\mu}^{3}\hat{g}_{3}+7{\mu}^{2}\hat{g}_{2}+{\mu}-\left(  N_{\textrm{cut}}+1\right)  \left(  {\mu}^{3}\hat{g}_{3}+3{\mu}^{2}\hat{g}_{2}+{\mu}-\sum\nolimits_{n=0}^{N_{\textrm{cut}}}\hat{p}_{n}n^{3}\right)   &  \geq\sum\nolimits_{n=0}^{N_{\textrm{cut}}}\hat{p}_{n}n^{4}
\end{align}

\section{Optimisation of the photon number yields}
\label{sec:appB}

In this section, we bound the photon number yields $y_n$ estimated through the decoy-state technique. As in the previous section, we only consider a specific problem, i.e., the minimisation of the single-photon yield $y_1$, which is the most relevant contribution to the key rate in Eq.~\eqref{eq:rate}. Other problems related to the maximisation of $y_1$ or to the optimisation of the other $y_n$'s, with $n \neq 1$, can be treated in a similar manner.

The objective function in the constrained-minimisation problem is $y_1$ and the constraints are given by the usual decoy-state equations:
\begin{align}
Y^{(u)} &  =\sum\nolimits_{n=0}^{\infty}\hat{p}_{n}^{(u)}\hat{y}_{n} \label{eq:Yu}\\
Y^{(v)} &  =\sum\nolimits_{n=0n}^{\infty}\hat{p}_{n}^{(v)}\hat{y}_{n} \label{eq:Yv}\\
Y^{(w)} &  =\sum\nolimits_{n=0}^{\infty}\hat{p}_{n}^{(w)}\hat{y}_{n} \label{eq:Yw}\\
0 &  \leq\hat{y}_{n}\leq1\\
\underline{p}_{n}^{(\mu)} &  \leq \hat{p}_{n}^{(\mu)} \leq \overline{p}_{n}^{(\mu)}
\end{align}
which should hold for all $n$ and all $\mu\in\{  u,v,w\}$. In the above equations, the hatted quantities are the unknowns. Moreover, we have indicated with $\underline{p}_{n}^{(\mu)}$ and $\overline{p}_{n}^{(\mu)}$ the lower and upper bounds, respectively, for the probability that an $n$-photon pulse is emitted by the light source when the mean photon number $\mu$ is prepared. These bounds are calculated from problems analogous to the one in Eq.~\eqref{ref_problem} for the probabilities $p_0^{(u)}$, $p_1^{(u)}$, $p_2^{(u)}$ and $p_3^{(u)}$. The lower (upper) bounds for the remaining probabilities for the signal state with intensity $u=0.42$ are conservatively assumed to be equal to 0 (to 1). The probabilities for the decoy state with intensity $v=0.02$ and for the vacuum state with intensity $w=10^{-4}$ are taken equal to the corresponding Poissonian distribution. This is justified by the fact that when the mean photon number is so small, we did not observe any practical deviation from a Poissonian distribution (see bottom-right diagram in Fig.~\ref{fig:fig6}, where the mean photon number is 0.1, i.e., 5 times bigger than the value considered here).

The constraints in Eqs.~\eqref{eq:Yu}-\eqref{eq:Yw} define an optimisation problem that is nonlinear in the unknowns. This obstacle can be overcome by replacing the unknowns $\hat{p}_n^{(\mu)}$ with their bounds $\underline{p}_{n}^{(\mu)}$ and $\overline{p}_{n}^{(\mu)}$, drawn from the optimisation problem in Eq.~\eqref{ref_problem}~\cite{CXC+14}. Moreover, the sums contain an infinite number of terms, which is not suitable for a practical scenario. This can be easily circumvented using the same technique demonstrated for the problem in Sec.~\ref{sec:appA} and in the literature~\cite{TCL16}, i.e., by cutting the sums to a finite value $N_{\textrm{cut}}$ and then bounding the residual. As a result, we obtain the following constraints:
\begin{align}
\overline{\Gamma}_{\mu}   &=      1-\sum\nolimits_{n=0}^{N_{\textrm{cut}}}\underline{p}_{n}^{(\mu)} \\
\overline{Y}^{(\mu)}      &\geq   \sum\nolimits_{n=0}^{N_{\textrm{cut}}}\underline{p}_{n}^{(\mu)}\hat{y}_{n}  \\
\underline{Y}^{(\mu)}     &\leq   \sum\nolimits_{n=0}^{N_{\textrm{cut}}}\overline{p}_{n}^{(\mu)}\hat{y}_{n} + \overline{\Gamma}_{\mu}\\
0 &  \leq\hat{y}_{n}\leq1.
\end{align}
where we have indicated the upper and lower bounds to the experimental yields for the mean photon number $\mu$ as $\overline{Y}^{(\mu)}$ and $\underline{Y}^{(\mu)}$, respectively, and the upper bound for the residual sum as $\overline{\Gamma}_{\mu}$.

\section{Optimisation of the photon number error rates}
\label{sec:appC}

The optimisation problem for the single-photon error rate $e_1$ follows similar lines as in the previous sections. In QKD, and in the key rate equation \eqref{eq:rate}, it is important to find an upper bound to the error rate. This can be achieved by replacing the bounds $\underline{Y}^{(\mu)}$ and $\overline{Y}^{(\mu)}$ in the previous section with the bounds for the bit error rate $\underline{B}^{(\mu)}=\underline{Y}^{(\mu)} \underline{E}^{(\mu)}$ and $\overline{B}^{(\mu)}=\overline{Y}^{(\mu)} \overline{E}^{(\mu)}$, respectively, and $\hat{y}_{n}$ with $\hat{b}_{n}$. Then, after maximising $\hat{b}_{1}$, $\overline{e}_{1}$ will be given by:
\begin{equation}
\overline{e}_{1}=\frac{\overline{b}_{1}}{\underline{y}_{1}},
\end{equation}
where $y_1$ has been obtained from the yield optimisation problem. After straightforward steps, we obtain an explicit upper bound for the single-photon error rate~\cite{LDY+12} which is the one used to draw the plots in Fig.~\ref{fig:fig7}:
\begin{equation}
\overline{e}_{1}  = \frac{ \overline{B}^{(u)} - \underline{p}_{0} \underline{y}_{0} \underline{e}_{0}}{\underline{p}_{1} \underline{y}_{1}}.
\label{eq_e1max}
\end{equation}

\end{widetext}


\end{document}